\PassOptionsToPackage{prologue,dvipsnames}{xcolor}
\documentclass[sigconf,divpsnames]{acmart}
\AtBeginDocument{%
  }

\copyrightyear{2025}
\acmYear{2025}
\setcopyright{rightsretained}
\acmDOI{10.1145/3711896.3737255}
\acmConference[KDD '25] {Proceedings of the 31st ACM SIGKDD Conference on Knowledge Discovery and Data Mining V.2}{August 3--7, 2025}{Toronto, ON, Canada.}
\acmBooktitle{Proceedings of the 31st ACM SIGKDD Conference on Knowledge Discovery and Data Mining V.2 (KDD '25), August 3--7, 2025, Toronto, ON, Canada}
\acmISBN{979-8-4007-1454-2/25/08}
\usepackage{lipsum}
\usepackage{enumitem}
\usepackage{amsmath}
\usepackage{multicol,multirow,array}
\usepackage{subcaption,adjustbox}
\newcommand{\ToolX}{MTMH}
\newcommand{\draft}[1]{\textcolor{black}{#1}}

\begin{document}

\title{Optimizing Recall or Relevance? A Multi-Task Multi-Head Approach for Item-to-Item Retrieval in Recommendation}
\author{Jiang Zhang}
\affiliation{%
  \institution{Meta Platforms, Inc.}
  \city{Menlo Park}
  \state{CA}
  \country{USA}
}
\email{jiangzhang2024@meta.com}
\author{Sumit Kumar}
\affiliation{%
  \institution{Meta Platforms, Inc.}
  \city{Seattle}
  \state{WA}
  \country{USA}
}
\email{sumitkumar@meta.com}
\author{Wei Chang}
\affiliation{%
  \institution{Meta Platforms, Inc.}
  \city{Menlo Park}
  \state{CA}
  \country{USA}
}
\email{mrweichang@meta.com}
\author{Yubo Wang}
\affiliation{%
  \institution{Meta Platforms, Inc.}
  \city{Bellevue}
  \state{WA}
  \country{USA}
}
\email{yubowang@meta.com}
\author{Feng Zhang}
\affiliation{%
  \institution{Meta Platforms, Inc.}
  \city{Austin}
  \state{TX}
  \country{USA}
}
\email{fengzhang1@meta.com}
\author{Weize Mao}
\affiliation{%
  \institution{Meta Platforms, Inc.}
  \city{Menlo Park}
  \state{CA}
  \country{USA}
}
\email{wzmao@meta.com}
\author{Hanchao Yu}
\affiliation{%
  \institution{Meta Platforms, Inc.}
  \city{Menlo Park}
  \state{CA}
  \country{USA}
}
\email{yhcece@gmail.com}
\author{Aashu Singh}
\affiliation{%
  \institution{Meta Platforms, Inc.}
  \city{Menlo Park}
  \state{CA}
  \country{USA}
}
\email{aashusingh@meta.com}
\author{Min Li}
\affiliation{%
  \institution{Meta Platforms, Inc.}
  \city{Menlo Park}
  \state{CA}
  \country{USA}
}
\email{minli@meta.com}
\author{Qifan Wang}
\affiliation{%
  \institution{Meta Platforms, Inc.}
  \city{Menlo Park}
  \state{CA}
  \country{USA}
}
\email{wqfcr@fb.com}
\renewcommand{\shortauthors}{Jiang Zhang et al.}

% , Sumit Kumar, Wei Chang, Yubo Wang, Feng Zhang, Weize Mao, Hanchao Yu \\ Aashu Singh, Min Li, Qifan Wang
\begin{abstract}
% background
% The task of item-to-item (I2I) retrieval is to identify a set of relevant and highly engaged items given a trigger item.
% It is a important component in modern recommendation systems, where items engaged by users in the past are used as trigger items to retrieve relevant items for future engagement. However, existing I2I retrieval models in industry are mainly build from the co-engagement data, over-emphasizing the co-engagement pattern and failing to optimize the semantic relevance, leading to overfitting short-term co-engagement data and sacrificing long-term values such as novel interest discovery and content diversity.
The task of item-to-item (I2I) retrieval is to identify a set of relevant and highly engaging items based on a given item. I2I retrieval is a crucial component in modern recommendation systems, where users' previously engaged items serve as trigger items to retrieve relevant content for future engagement. However, existing I2I models in industry are primarily built on co-engagement data and optimized using the recall measure, which overly emphasizes co-engagement patterns while failing to capture semantic relevance. This often leads to overfitting short-term co-engagement trends at the expense of long-term benefits such as discovering novel interests and promoting content diversity.
% % key challenge
% Optimizing I2I semantic relevance often comes at the cost of sacrificing co-engagement efficiency, as semantically relevant items may not always have high co-engagement rates.
% proposal
To address this challenge, we propose \textbf{\ToolX}, a \textbf{M}ulti-\textbf{T}ask and \textbf{M}ulti-\textbf{H}ead I2I retrieval model that achieves both high recall and semantic relevance. Our model consists of two key components: 1) a multi-task learning loss for formally optimizing the trade-off between recall and relevance, and 2) a multi-head I2I retrieval architecture for retrieving both highly co-engaged and semantically relevant items.
% results
We evaluate \ToolX~using proprietary data from a commercial platform serving billions of users and demonstrate that it can improve recall by up to 14.4\% and semantic relevance by up to 56.6\% compared with prior state-of-the-art models. We also conduct live experiments to verify that \ToolX~can enhance both short-term consumption metrics and long-term user-experience-related metrics.
Our work provides a principled approach for jointly optimizing I2I recall and semantic relevance, which has significant implications for improving the overall performance of recommendation systems.
\end{abstract}




\maketitle

\section{Introduction}
% 1. T2I task and problems
Item-to-item (I2I) retrieval refers to the task of retrieving relevant and highly engaged items for a given item (also named trigger item), which is an important component in modern recommendation systems. In I2I retrieval, items from users' past engagement history are used as trigger items to retrieve new items for future engagement, providing users with personalized experience \cite{deshpande2004item,xue2019deep,schnabel2020debiasing}.

\begin{figure}[t]
    \centering
    \includegraphics[width=0.7\linewidth]{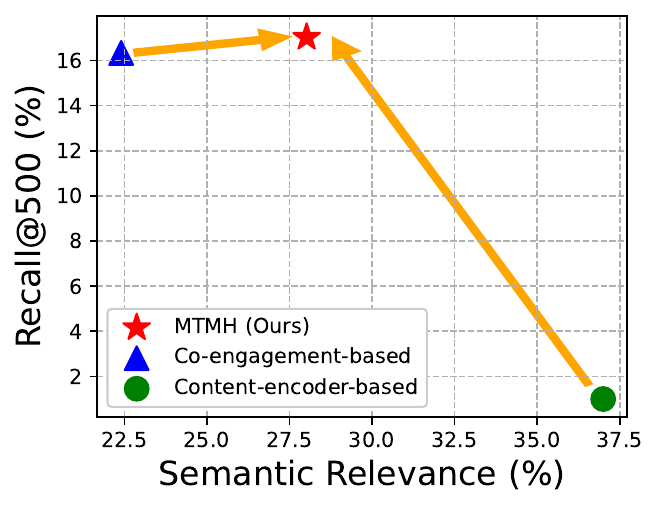}
    \vspace{-2mm}
    \caption{Recall vs. Relevance for I2I retrieval models. The co-engagement based model (\textcolor{blue}{blue triangle}) refers to an I2I model trained exclusively on co-engagement data, while the content-encoder based model (\textcolor{OliveGreen}{green circle}) represents an I2I model that directly utilizes embeddings generated by a pre-trained item content encoder.}
    \label{fig:intro}
    \vspace{-5mm}
\end{figure}

% 2. existing works over-emphasize engagement efficiency
% Most existing industry I2I models leverage co-engagement data to supervise their training, with the assumption that items users co-engaged within a short period are likely to be relevant. These I2I models often over-emphasize the co-engagement pattern during retrieval, since they are purely trained to optimize the recall metric from the co-engagement data without formally optimizing the semantic relevance (e.g. \cite{he2017neural,xue2019deep,schnabel2020debiasing,zhai2023revisiting,chen2024hllm}). This may cause I2I retrieval models overfitting the short-term I2I co-engagement data at the expense of sacrificing long-term values such as user retention, content diversity, and novel interest discovery \cite{wang2022surrogate}.
Most existing industry I2I models utilize co-engagement data to guide their training, operating under the assumption that items users engage with in quick succession are likely to be relevant. These I2I models often place excessive emphasis on co-engagement patterns during retrieval because they are primarily trained to optimize recall metrics based on this data, without explicitly optimizing for semantic relevance \cite{he2017neural,xue2019deep,schnabel2020debiasing,zhai2023revisiting,chen2024hllm}. This focus can lead to I2I retrieval models overfitting to short-term co-engagement data, potentially compromising long-term objectives such as user retention, content diversity, and the discovery of new interests \cite{wang2022surrogate}.

% 3. why and challenges
% Improving I2I semantic relevance is essential for enhancing the overall performance of recommendation systems and offers several key benefits. First, it can improve the user interest recall of retrieved items and provide more personalized user experience\cite{lv2019interest}. Second, it helps to surface fresh content more effectively, especially for cold content without enough user engagement data \cite{hidasi2013context}.  Finally, it mitigates the short-term co-engagement bias and creates a more healthy and valuable feedback loop with better long-term value in recommendation systems \cite{jadidinejad2020using}.
% However, optimizing I2I semantic relevance is non-trivial, since it is challenging to quantify the semantic relevance between two items, especially for multi-modal content (e.g. short videos). This says, there is a lack of explicit supervision labels for optimizing I2I semantic relevance. More fundamentally, there exists a trade-off between recall and semantic relevance in I2I retrieval: \textit{semantically relevant items may not always have high co-engagement rates}.
Enhancing the semantic relevance of I2I models is crucial for improving the overall performance of recommendation systems and offers several significant benefits. First, it can enhance the recall of user interests in retrieved items, thereby providing a more personalized user experience \cite{lv2019interest}. Second, it facilitates the effective surfacing of fresh contents, particularly for new contents that lacks sufficient user engagement data \cite{hidasi2013context}. Lastly, it helps mitigate short-term co-engagement bias, fostering a healthier and more valuable feedback loop that contributes to better long-term outcomes in recommendation systems \cite{jadidinejad2020using}.
However, optimizing I2I semantic relevance is challenging, primarily because quantifying the semantic relevance between two items is difficult, especially for multi-modal content such as short videos. This challenge is compounded by the lack of explicit supervision labels for optimizing semantic relevance. Fundamentally, there is a trade-off between recall and semantic relevance in I2I retrieval: \textit{semantically relevant items may not always exhibit high co-engagement rates}.

% 4. Preliminary results
% To illustrate this trade-off, we conduct a comparison of the recall and relevance of two I2I retrieval mdoels: 1) a co-engagement based model purely trained on co-engagement data; and 2) a content-encoder based model that leverages a large pre-trained content encoder to generate item embeddings.
% %
% As shown in Figure \ref{fig:intro}, the co-engagement based model achieves high recall but poor semantic relevance. In contrast, the content-encoder based model is capable of retrieving items with high semantic relevance. However, its recall is less than 1\%, significantly worse compared to the co-engagement based model. This trade-off highlights the challenge of balancing co-engagement rates and semantic relevance in I2I retrieval models (see Section \ref{sec:prelim} for experimental details).
To illustrate this trade-off, we compare the recall and relevance of two I2I retrieval models: (1) a co-engagement based model trained solely on co-engagement data, and (2) a content-encoder based model that utilizes a large pre-trained content encoder to generate item embeddings.
As depicted in Figure \ref{fig:intro}, the co-engagement based model achieves high recall but exhibits poor semantic relevance. In contrast, the content-encoder based model is adept at retrieving items with high semantic relevance, yet its recall is less than 1\%, significantly lower than that of the co-engagement based model. This trade-off underscores the challenge of balancing co-engagement rates and semantic relevance in I2I retrieval models. More experimental details are provided in Section \ref{sec:prelim}.

% 5. proposals
% To address the above challenges, in this work, we propose \textbf{\ToolX}, a \textbf{M}ulti-\textbf{T}ask and \textbf{M}ulti-\textbf{H}ead I2I retreival model, which achieves optimal balance between recall and semantic relevance, as shown in Figure \ref{fig:intro}. The design of \ToolX~contains two key components: 1) a multi-task learning loss for jointly optimizing the recall and semantic relevance, and 2) a multi-head I2I model architecture for retrieving both highly co-engaged and semantically relevance items.
% Specifically, the co-engagement loss is designed to maximize the I2I co-engagement rate, while the semantic relevance loss is designed to distill the semantic information of items from a pre-traiend content encoder into the learnt item embeddings. The multi-task learning loss is computated as a weighed sum of engagement loss and relevance loss, providing a principled approach to jointly optimize I2I co-engagement efficiency and semantic relevance during training (see Section \ref{subsec:relevance}). Moreover, we design a multi-head I2I retrieval model that contains an engagement head and a relevance head. The engagement head is trained purely based on engagement loss to select highly co-engaged items, while the relevance head is trained via multi-task learning loss to select highly relevant and co-engaged items. By merging candidates from both heads, \ToolX~ is able to achieve both improved I2I recall and semantic relevance during serving time (see Section \ref{subsec:serving}).
To address the aforementioned challenges, we propose \textbf{\ToolX}, a \textbf{M}ulti-\textbf{T}ask and \textbf{M}ulti-\textbf{H}ead I2I retrieval model that achieves an optimal balance between recall and semantic relevance, as illustrated in Figure \ref{fig:intro}. The design of \ToolX~incorporates two key components: (1) a multi-task learning loss for jointly optimizing recall and semantic relevance, and (2) a multi-head I2I model architecture for retrieving items that are both highly co-engaged and semantically relevant.
Specifically, the co-engagement loss is crafted to maximize the co-engagement rate (or recall), while the semantic relevance loss is designed to preserve the semantic similarity between items by distilling item semantic knowledge from a pre-trained content encoder into the learned item embeddings. The multi-task loss is computed as a weighted sum of the engagement loss and relevance loss, providing a principled approach to jointly optimize co-engagement efficiency and semantic relevance during training. Moreover, we design a multi-head I2I retrieval model that includes an engagement head and a relevance head. The engagement head is trained solely on engagement loss to select highly co-engaged items, while the relevance head is trained using the multi-task learning loss to select items that are both highly relevant and co-engaged. By merging retrieved items from both heads, \ToolX~achieves improved recall and semantic relevance simultaneously during serving time.

% 6. contributions
We evaluate the performance of \ToolX~using proprietary data from a commercial platform serving billions of users. Our results show that \ToolX~can improve I2I retrieval recall by up to 14.4\% and semantic relevance by up to 56.6\%, achieving the best trade-off between these two metrics compared with all baselines (see Section \ref{subsec:main_results}).
To further validate the effectiveness of \ToolX, we deploy \ToolX~ on this commercial platform and conduct online A/B testing. Our online evaluation results demonstrate that \ToolX~not only successfully improves consumption metrics such as daily active users and time spent, but also significantly enhances user-experience-related metrics such as user interest recall, novel interest discovery rate, content diversity and freshness (see Section \ref{subsec:online}).
We summarize our key contributions as follows:
\vspace{-3mm}
\begin{itemize}[leftmargin=*]
    \item We systematically examine the fundamental trade-off between recall and semantic relevance in I2I retrieval, uncovering their interconnections and highlighting the challenges of balancing and optimizing these two metrics.
    \item We propose \ToolX, a \textbf{M}ulti-\textbf{T}ask and \textbf{M}ulti-\textbf{H}ead I2I retrieval model, which provides a principled approach for jointly optimizing the trade-off between I2I co-engagement rate and semantic relevance.
    \item We evaluate \ToolX~on proprietary data from a commercial platform serving billions of users and demonstrate that it can improve recall by up to 14.4\% and semantic relevance by up to 56.6\% compared with prior SOTAs.
    \item We integrate \ToolX~into production to verify that \ToolX~can increase both topline consumption metrics and long-term user-experience-related metrics.
\end{itemize}

% \noindent\textbf{Paper Organization:} The rest of the paper is organized
% as follows. Section \ref{sec:prelim} shows some preliminary results on the recall and semantic relevance trade-off in I2I retrieval, which motivates the design of \ToolX. Section \ref{sec:method} describes the design of \ToolX. We present the evaluation results and experimental findings in Section \ref{sec:results}. Section \ref{sec:related} summarizes prior literature before concluding with Section \ref{sec:conclusion}.

\section{Preliminary Study}
\label{sec:prelim}

\begin{figure}[t]
    \centering
    \begin{subfigure}[t]{0.49\linewidth}
        \includegraphics[width=\linewidth]{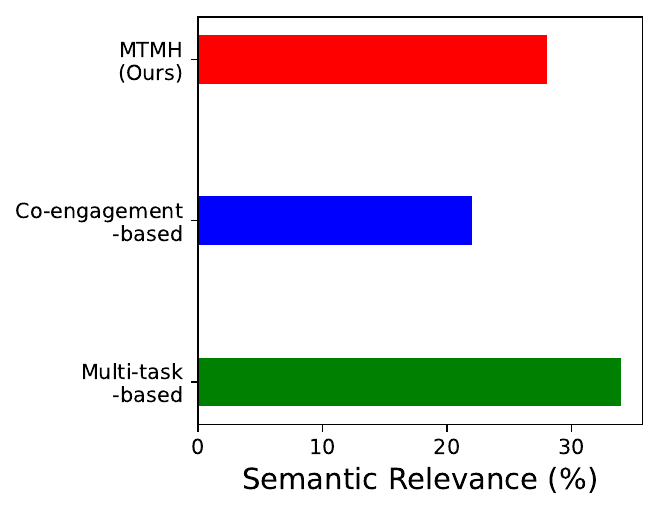}
        \caption{Semantic relevance.}
        \label{fig:intro2-relevance}
    \end{subfigure}
    \begin{subfigure}[t]{0.49\linewidth}
        \includegraphics[width=\linewidth]{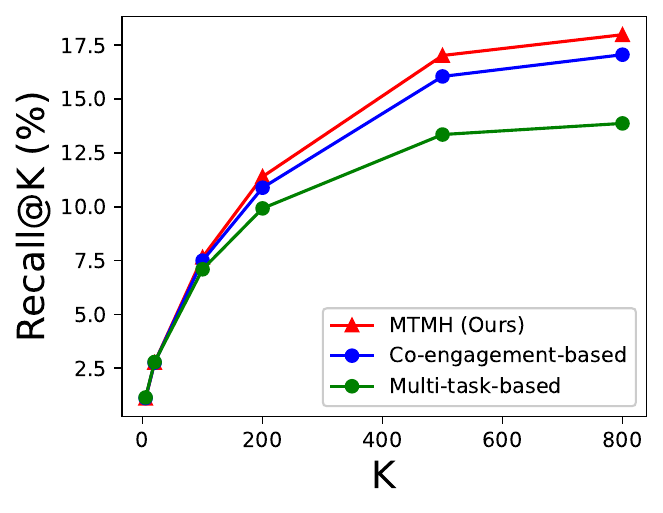}
        \caption{Recall with varying $K$.}
        \label{fig:intro2-recall}
    \end{subfigure}
    \begin{subfigure}[t]{0.97\linewidth}
        \includegraphics[width=\linewidth]{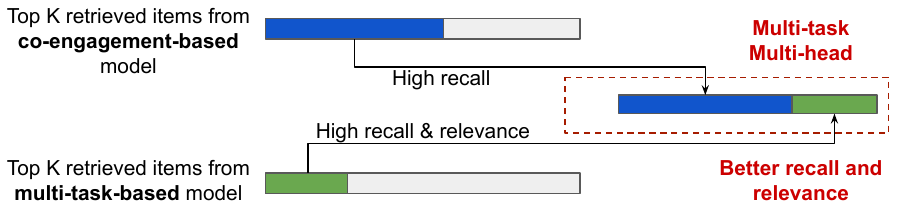}
        \caption{How to achieve both better recall and relevance.}
        \label{fig:intro2-mtmh}
    \end{subfigure}
    \vspace{-2mm}
    \caption{Recall vs relevance for I2I retrieval models. The co-engagement based model (\textcolor{blue}{blue line}) is trained via co-engagement loss only, while the multi-task-based model (\textcolor{OliveGreen}{green line}) is trained on multi-task learning loss (co-engagement loss + relevance loss).}
    \label{fig:intro2}
    \vspace{-6mm}
\end{figure}

In this section, we conduct a preliminary study to investigate the trade-off between recall and semantic relevance in I2I retrieval, providing insights and motivating the \ToolX's design.

% General setup
\noindent\textbf{Experimental setup.} We train the aforementioned I2I retrieval models using user data from period $T_1$ and evaluate their performance using data from period $T_2$ after $T_1$. During evaluation, we first sample a set of users, and for each user we select 50 engaged items from their past user interaction history (UIH) as trigger items. Next, for each trigger item, we conduct Approximate Nearest Neighbor (ANN) search based on embedding cosine similarity to identify the top 2000 nearest candidate items and employ a preranker model to further select top 30 out of 2000 candidates for each trigger item. Finally, for each user, we sort candidate items retrieved by 50 trigger items based on their ranking scores and select top 500 candidate items to measure recall and semantic relevance. Note that the recall@500 is computed as the percentage of ground-truth engaged items from future UIH hit by these 500 retrieved items. Semantic relevance is measured by the average topic category (from human labels) match rates between each trigger item and each candidate item retrieved by this trigger.

% Key points:
% The insights gained from these experiments motivate the design motivates our design of \ToolX.
% 1. Why not using LLM? Content feature as input -> worse recall but higher relevance
% 2. Why using MTMH? Show the saturation issue of recall for relevance head: hit rate stops increasing for larger K.

Next, we present our key findings from this preliminary study.
We start with the following question:\\
\noindent\textbf{Q1: Can the embeddings generated by a large pre-trained content encoder be directly used for I2I retrieval?} To answer this question, we compare the recall and semantic relevance performance of the following two baseline models: \textit{1) Co-engagement based model}: This model is trained on co-engagement data to learn item embeddings. The training objective is to maximize the I2I co-engagement rate. \textit{2) Content-encoder based model}: This model leverages a large pre-trained content encoder with superior content understanding capability to generate content embeddings for items, and directly uses them for I2I retrieval.

\noindent\textbf{Results for Q1.} Figure \ref{fig:intro} presents the evaluation results of co-engagement based and content-encoder based models. We observe that the co-engagement based model achieves 22.4\% semantic relevance, while the content-encoder based model achieves 37.5\%. This indicates that item embeddings generated by the pre-trained content encoder exhibit significantly better semantic relevance compared to the co-engagement based model. However, the recall of the content-encoder based model is less than 1\%, while the co-engagement based model achieves 16.3\%. This stark contrast highlights a strong disconnection between semantic relevance and co-engagement rate: \textit{high semantic relevance does not necessarily translate into a high recall in I2I retrieval}. Motivated by this, we introduce a multi-task learning loss that jointly optimizes co-engagement rate and semantic relevance. %This approach enables the model to generate item embeddings that effectively capture both co-engagement and semantic information, resulting in improved recall and semantic relevance (see \ToolX~in Figure \ref{fig:intro}).

% This says, the item embeddings generated by the pre-trained content encoder have significantly better I2I semantic relevance compared with the co-engagement-based model. However, the recall of content-encoder-based model is less than 1\% versus 16.3\% for co-engagement-based model. This clearly shows that there is a strong disconnect between semantic relevance and co-engagement rate: \textit{high semantic relevance does not necessary translate into high co-engagement rate in I2I retrieval}. Motivated by this, in Section \ref{subsec:relevance}, we propose a multi-task learning loss for I2I retrieval model that optimizes both I2I co-engagement rate and semantic relevance in \ToolX. This approach enables the item embeddings generated by \ToolX~ to effectively capture both co-engagement information and semantic information, achieving both higher recall and semantic relevance (see \ToolX~in Figure \ref{fig:intro}).

\noindent\textbf{Q2: Is it possible to simultaneously enhance recall and semantic relevance of I2I retrieval model?}
Although multi-task learning provides a structured approach to optimizing the trade-off between recall and semantic relevance, it does not fully resolve the fundamental challenge of balancing these two objectives: \textit{maximizing recall and maximizing semantic relevance cannot be achieved simultaneously in multi-objective optimization}. We hypothesize that items co-engaged by a user are not always semantically relevant, meaning that retrieving only high semantic relevance items may overlook highly engaged but less relevant items (e.g., those popular items). To test this, we compare the recall and semantic relevance of a model trained via multi-task learning (denoted as the multi-task based model) with a co-engagement based model. The formal definition of the multi-task learning is provided in Section \ref{subsec:relevance}.

\noindent\textbf{Results for Q2.}
As shown in Figure \ref{fig:intro2-relevance}, the multi-task based model achieves over a 50\% increase in semantic relevance compared to the co-engagement based model, which is expected due to its multi-task relevance modeling. In contrast, as illustrated in Figure \ref{fig:intro2-recall}, the recall of the multi-task based model is lower than that of the co-engagement-based model. Specifically, when $K$ is smaller than 200, the multi-task based model achieves a recall comparable to that of the co-engagement based model, suggesting that \textit{highly relevant candidates can also exhibit high engagement efficiency}. However, as $K$ increases, the recall of the multi-task based model begins to plateau much earlier than that of the co-engagement based model. This finding indicates that \textit{retrieving more semantically relevant items does not necessarily increase the overall recall}.
% As shown in Figure \ref{fig:intro2-relevance}, the multi-task based model achieves more than 50\% increase in semantic relevance compared to the co-engagement based model, which is expected dueo to its multi-task relevance modeling. In contrast, as illustrated in Figure \ref{fig:intro2-recall}, the recall of the multi-task based model is worse than the co-engagement based model. Specifically, we observe that when $K$ is smaller than 200, multi-task based model can actually achieve a comparable recall as the co-engagement based model. This implies that \textit{highly relevant candidates can also have high engagement efficiency}. However, when $K$ becomes larger, the recall of multi-task based model starts to saturate, much earlier than the recall of co-engagement based model. This finding indicates that \textit{the increased amount of semantically relevant items retrieved does not cover more ground-truth items}.

Motivated by these findings, we design a multi-task multi-head retrieval architecture. In this setup, one head is trained to retrieve highly engaged items by minimizing only the co-engagement loss, while the other head is trained to retrieve highly relevant items by minimizing the multi-task learning loss (see Section \ref{subsec:serving} for details). By merging candidates retrieved from both heads during serving (as illustrated in Figure \ref{fig:intro2-mtmh}), \ToolX~ is able to retrieve items that are extremely high in co-engagement (albeit with less semantic relevance) from the first head, and items that are highly semantically relevant from the second head, leading to both higher recall and enhanced semantic relevance.
% Motivated by the above findings, we design a multi-task multi-head retrieval architecture, where one head is trained to retrieve highly engaged items by solely minimizing the co-engagement loss, and the other head is trained to retrieve highly relevant items by minimizing multi-task learning loss (see Section \ref{subsec:serving} for details). By merging candidates retrieved from both heads during serving (see Figure \ref{fig:intro2-mtmh}), \ToolX~ can retrieve both extremely high co-engaged items (with less semantic relevance) from head one, and highly semantic relevant items from head two, resulting in both higher recall and semantic relevance.

\section{Methodology}
% In this section, we present the design of \ToolX, which consists of three key components: 1) a pre-trained content encoder for converting multi-modal item content into semantic embeddings (see Section \ref{subsec:encoder}); 2) a multi-task learning loss for jointly optimizing I2I recall and semantic relevance (see Section \ref{subsec:relevance}) and 3) a multi-head I2I model architecture for retrieving both highly co-engaged and semantically relevant items (see Section \ref{subsec:model} and \ref{subsec:serving}).
% \label{sec:method}

\begin{figure*}[t]
    \centering
    \begin{subfigure}{0.18\linewidth}
        \includegraphics[width=\linewidth]{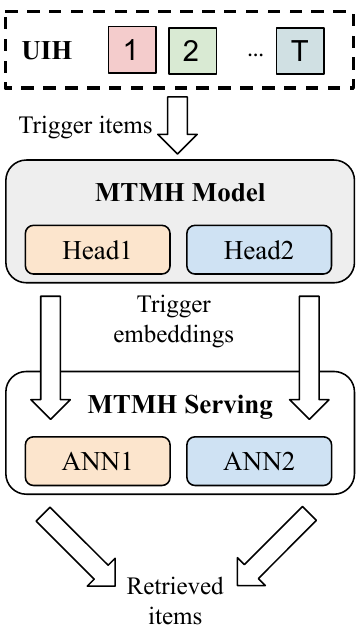}
        \caption{\ToolX~ workflow.}
        \label{fig:design-overview}
    \end{subfigure}
    \begin{subfigure}{0.815\linewidth}
        \includegraphics[width=\linewidth]{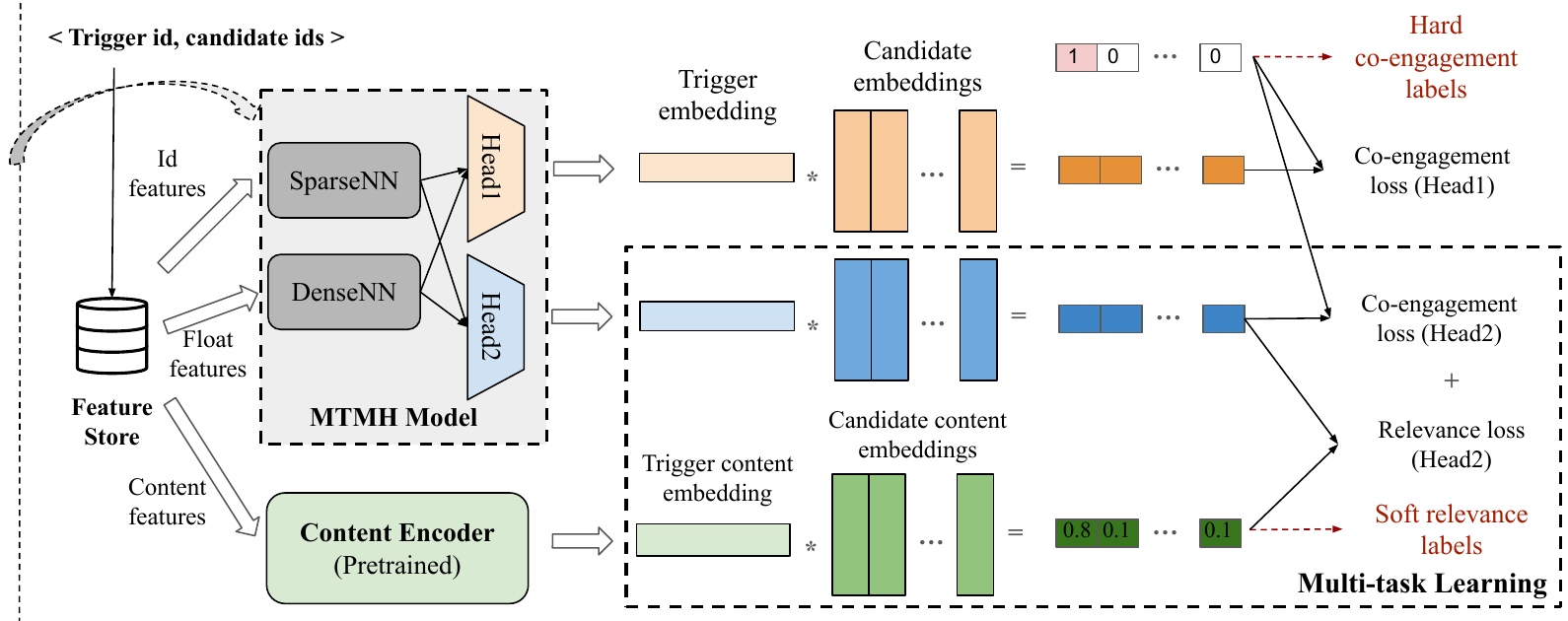}
        \caption{\ToolX~ model and training loss.}
        \label{fig:design-loss}
    \end{subfigure}
    \caption{Overview of the proposed \ToolX~ approach. 1) The multi-task learning consists of a co-engagement loss and a relevance loss, where the co-engagement loss has the format of InfoNCE \cite{oord2018representation}, with the hard labels from the co-engagement data and the relevance loss is based on knowledge distillation from the pre-trained content encoder. 2) The multi-head design includes an engagement head (w/ only co-engagement loss) and a relevance head (w/ multi-task loss). 3) The pre-trained content encoder is used to generate relevance supervision for transferring the content semantic knowledge to the learned embeddings.}
    \label{fig:main}
    \vspace{-3mm}
\end{figure*}

% \begin{figure}[t]
%     \centering
%     \includegraphics[width=0.97\linewidth]{docs/figs/kdd-overview.pdf}
%     \caption{Overview of the proposed \ToolX~ approach.}
%     \label{fig:design-overview}
%     \vspace{-5mm}
% \end{figure}

\subsection{Overview of \ToolX~Approach}
We first provide an overview of our \ToolX~model in figure \ref{fig:main}. The goal of \ToolX~ is to learn the item embeddings and use them to retrieve highly engaged and relevant items based on past user engagements. \ToolX~ essentially consists of three key components: 1) a multi-task learning module for jointly optimizing recall and semantic relevance (Section \ref{subsec:relevance}); 2) a multi-head design for retrieving both highly co-engaged and semantically relevant items (Section \ref{subsec:model}) and 3) a pre-trained content encoder used to distill multi-modal content knowledge into the learned item embeddings (Section \ref{subsec:encoder});
\label{sec:method}
% past user interaction history (UIH) as \textit{trigger items} to identify a  set of the most similar \textit{candidate items} from the entire item universe for future user engagement. It represents items as embeddings based on their features, and it is trained to  maximize the likelihood of selecting the co-engaged candidate items from the candidate item universe based on embedding similarity. 

% To train the \ToolX~ model, we need to collect a set of positive and negative <\textit{trigger, candidate}> item pairs. Specifically, we construct positive <\textit{trigger, candidate}> item pairs from user interaction history (UIH). Suppose a UIH contains $T$ past engaged items in chronological order, we pair the $T$-th engaged items with the previous $T$-$1$ engaged items in this UIH to construct $T$-$1$ positive item pairs. To enhance the relevance of these positive pairs, we apply the Search-based Interest Model (SIM \cite{pi2020search,Trinity}) to identify semantically relevant items from the user's UIH, and assign larger weights to these semantically relevant item pairs. Additionally, we randomly sample a set of items that are not engaged by the user and pair them with the engaged items to construct negative item pairs for each UIH.
To train the \ToolX~ model, we first collect a set of positive and negative <\textit{trigger, candidate}> item pairs. Specifically, we construct positive pairs from user interaction history (UIH). Given a UIH containing $T$ past engaged items in chronological order, we pair the $T$-th item with each of the previous $T$-$1$ items, forming $T$-$1$ positive pairs. To further enhance the relevance of these positive pairs, we leverage the Search-based Interest Model (SIM \cite{pi2020search,Trinity}) to identify semantically relevant items within the user's UIH and assign higher weights. Additionally, we construct negative item pairs by randomly sampling items that the user has not engaged with and pairing them with the engaged items in each UIH.

\subsection{Multi-task Learning}
\label{subsec:relevance}

% \begin{figure*}
%     \centering
%     \includegraphics[width=0.8\linewidth]{docs/figs/kdd-loss.pdf}
%     \caption{Multi-task learning module consists of a co-engagement loss and a relevance loss. The co-engagement loss has the format of InfoNCE \cite{oord2018representation}, with the hard labels from the co-engagement data. The relevance loss is based on knowledge distillation, with the soft semantic relevance labels computed from the content embedding generated by the pre-trained content encoder.}
%     \label{fig:design-loss}
%     \vspace{-3mm}
% \end{figure*}

In this subsection, we present the multi-task learning module which optimizes both recall and semantic relevance in I2I retrieval model. At a high level, our multi-task learning loss is computed as the weighted sum of a co-engagement loss and a semantic relevance loss, as shown in Figure \ref{fig:design-loss}. The co-engagement loss is designed to maximize the embedding similarity between positive item pairs (i.e. co-engaged items) while minimize the embedding similarity between negative item pairs. The semantic relevance loss is designed to minimize the contrastive semantic information loss between content embeddings generated by the content encoder and item embeddings learned by I2I retrieval model.

\noindent\textbf{Co-engagement loss.} We employ the widely used InfoNCE loss \cite{oord2018representation} as the co-engagement loss, which is formally defined as:
\begin{equation}
\begin{split}
\small
\label{eq:engage-loss}
    L_{e} = -\sum_{i=1}^{M}\sum_{j=1}^{N}\log p_{i,j}^{+}, \ \ \ \ \ \ \ \ \ \ \\
% \end{equation}
% %where 
% \begin{equation}
    p_{i,j}^{+}= \frac{e^{<E_i, E_{i,j}^{+}>}}{e^{<E_i, E_{i,j}^{+}>} + \sum_{k=1}^{k=L}e^{<E_i, E_{i,k}^{-}>}}
\end{split}
\end{equation}
where $p_{i,j}^{+}$ is the predicted probability for trigger item $i$ to identify positive candidate item $j$ from a set of items based on item embedding similarity. 
$E_i$ is the $i$-th trigger item embedding. $E_{i,j}^{+}$ is the $j$-th item embedding which is positively paired with trigger item $i$, and $E_{i,k}^{-}$ is the $k$-th item embedding which is negatively paired with trigger item $i$. $M$ is the total number of trigger items, whereas $N$ is the number of candidate items which are positively paired with each trigger item. $L$ is the number of items which are negatively paired with each trigger item, and $<x,y>$ represents the dot product between embedding $x$ and $y$. By minimizing the co-engagement loss, the embeddings of positive item pairs are pulled closer, while the embeddings of negative item pairs are pushed apart.

\noindent\textbf{Semantic relevance loss.} Co-engagement modeling does not guarantee the preservation of semantic relevance in the learned item embeddings, often resulting in suboptimal item relevance. To address this, we introduce a semantic relevance loss that aligns the contrastive similarity between item embeddings generated by the I2I retrieval model with the contrastive similarity between item content embeddings produced by the content encoder. In this way, the semantic content knowledge is distilled into the learned embeddings. We use the content embeddings generated by the pre-trained content encoder to build soft semantic relevance labels between <trigger, candidate> pairs, and then use these soft relevance labels to guide the semantic relevance optimization (see Figure \ref{fig:design-loss} for details). This process is also known as knowledge distillation \cite{hinton2015distilling}. Formally, the relevance loss is defined as:
\begin{equation}
\label{eq:relev-loss}
\small
L_r = \sum_{i=1}^{M}\sum_{j=1}^{N} D_{KL}(Q_{i,j}||P_{i,j}) = \sum_{i=1}^{M}\sum_{j=1}^{N}\big(q_{i,j}^{+}\log\frac{q_{i,j}^{+}}{p_{i,j}^{+}} + \sum_{k=1}^{L}q_{i,k}^{-}\log\frac{q_{i,k}^{-}}{p_{i,k}^{-}}\big)
\end{equation}
where:
\[
q_{i,j}^{+} = \frac{e^{<F_i, F_{i,j}^{+}>}}{e^{<F_i, F_{i,j}^{+}>} + \sum_{k=1}^{k=L}e^{<F_i, F_{i,k}^{-}>}}, \
q_{i,k}^{-} = \frac{e^{<F_i, F_{i,k}^{-}>}}{e^{<F_i, F_{i,j}^{+}>} + \sum_{k=1}^{k=L}e^{<F_i, F_{i,k}^{-}>}}
\]
\[
p_{i,k}^{-} = \frac{e^{<E_i, E_{i,k}^{-}>}}{e^{<E_i, E_{i,j}^{+}>} + \sum_{k=1}^{k=L}e^{<E_i, E_{i,k}^{-}>}}
\]
$Q_{i,j}=[q_{i,j}^{+}, q_{i,1}^{-},...,q_{i,L}^{-}]$, and $P_{i,j}=[p_{i,j}^{+}, p_{i,1}^{-},...,p_{i,L}^{-}]$. $D_{KL}(Q_{i,j}||P_{i,j})$ denotes the KL divergence between probability distributions $Q_{i,j}$ and $P_{i,j}$. $F_i$ is the $i$-th trigger item content embedding, $F_{i,j}^{+}$ is the $j$-th item content embedding positively paired with trigger $i$, $F_{i,k}^{-}$ is the $k$-th item content embedding negatively paired with trigger $i$.

Note that $q_{i,j}^{+}$ and $q_{i,k}^{-}$ can be interpreted as the target probability for trigger item $i$ to identify positively paired item $j$ and negatively paired item $k$ based on item content embedding similarity respectively. $p_{i,j}^{+}$ (defined in Eq. (\ref{eq:engage-loss})) and $p_{i,k}^{-}$ can be interpreted as the predicted probability for trigger item $i$ to identify positively paired item $j$ and negatively paired item $k$ based on item embedding similarity. By minimizing the KL divergence between $Q_{i,j}$ and $P_{i,j}$, the relative similarities between learned item embeddings are aligned with those of the item content embeddings generated by the content encoder. Hence, item embeddings generated by the I2I retrieval model can effectively preserve item semantic relevance.

\noindent\textbf{Multi-task loss.}  The multi-task learning loss is defined as:
\begin{equation}
    L_{mt} = L_e + w_r*L_r
    \label{eq:loss_mt}
\end{equation}
where $w_r$ is an hyper-parameter to control the weight of relevance loss. Increasing $w_r$ is expected to improve the semantic relevance of the model but might decrease the co-engagement rate (or recall) at the same time (see Section \ref{subsec:tradeoff} for details). Note that the default value of $w_r$ is 0.5, unless otherwise specified.

\begin{figure}[t]
    \centering
    \includegraphics[width=0.95\linewidth]{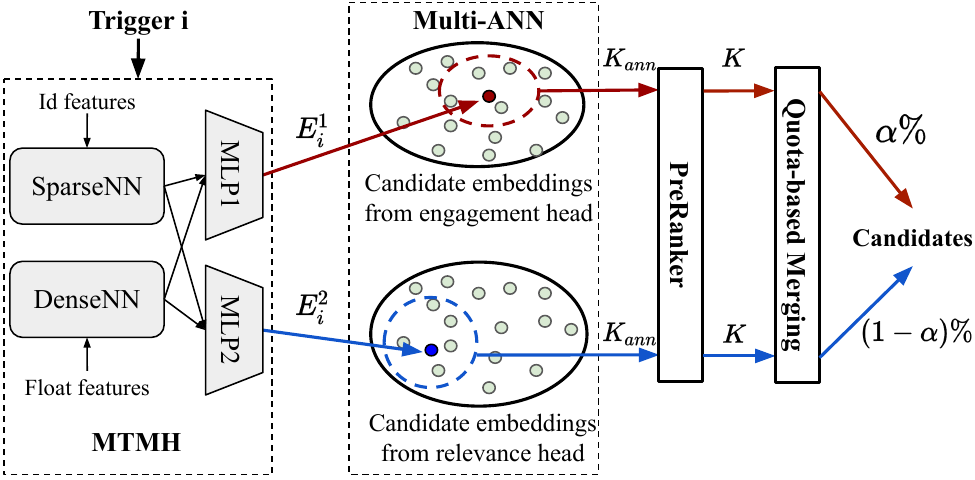}
    \caption{\ToolX~ serving pipeline contains three modules: 1) a multi-ANN module to retrieve $K_{ann}$ nearest candidates from each head; 2) a preranker module to select top $K$ out of $K_{ann}$ candidates from each head; 3) a quota-based merging module to merge candidates from the two heads.}
    \label{fig:design-serving}
    \vspace{-5mm}
\end{figure}
\subsection{Multi-head Architecture}
\label{subsec:model}
As discussed in Section \ref{sec:prelim}, although the I2I retrieval model trained via multi-task learning can retrieve items with high semantic relevance, it may miss some highly engaged but less relevant items. To overcome this challenge, we design a multi-head model with an engagement head and a relevance head.
Specifically, the engagement head is trained to minimize co-engagement loss $L_e$ (Eq. \ref{eq:engage-loss}) only, focusing on retrieving highly co-engaged items. In contrast, the relevance head is trained to minimize the multi-task loss $L_{mt}$ (Eq. \ref{eq:loss_mt}) to retrieve items with high semantic relevance. 

As demonstrated in Figure \ref{fig:design-serving}, at the bottom of the multi-head model, a Sparse Neural Network (SparseNN) is used to map each id feature of the input item into a unique embedding vector, and a Dense Neural Network (DenseNN) is used to map the float features of the input item into a latent vector. These vectors are then concatenated and fed into two MLP heads, each of which has the same input vectors but uses a dedicated MLP to map them into a different item embedding space. Notably, increasing the number of heads only slightly increases the total number of model parameters, since the majority of parameters come from SparseNN.

\subsection{Content Encoder}
\label{subsec:encoder}
\begin{figure}[t]
    \centering
    \includegraphics[width=\linewidth]{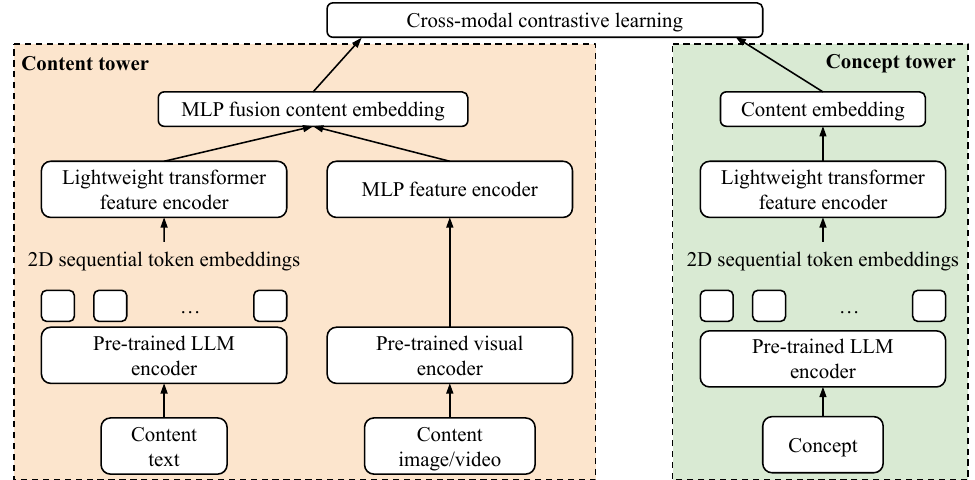}
    \vspace{-5mm}
    \caption{Model architecture of the content encoder. The content tower processes content inputs to generate multimodal embeddings, while the concept tower extracts semantic embeddings from text concepts. Contrastive loss optimizes relevance between content and concept representations.}
    \label{fig:uir}
    \vspace{-5mm}
\end{figure}

To effectively capture the item semantic representation, inspired by recent works LLM2VEC~\cite{behnamghader2024llm2vec} and VLM2VEC~\cite{jiang2024vlm2vec}, we propose a multimodal contrastive learning framework to learn the correspondence between content and concepts. Figure \ref{fig:uir} shows the model architecture. \draft{The content tower is composed of a pre-trained LLM text encoder \cite{touvron2023llama} and visual encoder \cite{xu2023demystifying} that extract multimodal representations from the content}. These representations are then aggregated through a Multi-layer Perceptron (MLP) module to obtain a single embedding. Similarly, the concept tower generates concept embeddings from a pre-trained LLM encoder. Contrastive loss between the content and concept embeddings is used to train this model. The pairs of content and concepts are sourced from multiple places, including user-added hashtags, user search queries, and multimodal LLM-generated tags. The relevance knowledge in this content encoder is then distilled to the learned \ToolX~ model. \draft{Note that we use VIT-H14-632M as the visual encoder model, and XLM-R-Large-550M as the text encoder model. Both the visual and text encoders have output dimension of 1024, and the final output content embedding output by fusion MLP has dimension 128 (see Table \ref{tab:model_hyperparameters} for details).}

\subsection{\ToolX~ Serving Strategy}
\label{subsec:serving}
Building on the top of \ToolX~ architecture, we now describe the serving strategy of \ToolX. As illustrated in Figure \ref{fig:design-serving}, the serving pipeline of \ToolX~ consists of three key modules: 1) a multi-ANN module to conduct ANN search and retrieve top $K_{ann}$ nearest candidates based on embeddings generated by each head in parallel; 2) a per-head preranker module to rank the candidates retrieved from each head and preserve top $K$ candidates for each head; and 3) a quota-based candidate merging module which merges top K candidates from each head based on their percentage quota.

\noindent\textbf{Multi-ANN module.} For candidate item embeddings generated by each head of \ToolX, we perform K-means clustering to divide items into different clusters offline. During online serving time, given a trigger item embedding generated by head $i$, we first select $C$ nearest clusters based on the embedding similarity between trigger embedding and cluster centroid embeddings. Then, we find the top $K_{ann}$ nearest candidates from items among these $C$ nearest clusters. Since there is no computation dependency between the ANN search for each head, \draft{the multi-ANN module can run per-head ANN search in parallel efficiently with limited overhead}. Note that in total, this module retrieves $2K_{ann}$ candidate items for each trigger item ($K_{ann}=O(1000)$ in \ToolX).

\noindent\textbf{Preranker module.} The module uses a multi-task user-to-item (U2I) model as preranker, in order to select top $K$ out of $2K_{ann}$ items retrieved by the multi-ANN module. These $K$ items are expected to have the highest probability of being engaged by the user. Since preranker is trained to maximize the likelihood of selecting items engaged by the user, it may be biased towards highly engaged items by users and thus assign low ranking scores to semantically relevant items. To mitigate such bias, we rank the $K_{ann}$ candidate items retrieved by each head separately, in order to guarantee that $K$ candidate items from each head are preserved ($K=O(10)$).

\noindent\textbf{Candidate merging module.} This module takes the output items of the preranker module as input, which is designed to merge the top $K$ candidate items retrieved from each head based on their percentage quota. Specifically, it selects the top $\alpha\%$ candidate items from the engagement head and the top $(100-\alpha)\%$ candidate items from the relevance head. Note that we remove duplicated candidates from each head to guarantee that the total amount of candidate items after merging is $K$. A larger $\alpha$ increases the number of candidates retrieved from the engagement head, potentially increasing I2I recall while decreasing I2I semantic relevance; conversely, a smaller $\alpha$ has the opposite effect. It is worth noting that $\alpha$ is a hyperparameter that can be adjusted during serving time. This means that we can flexibly trade between recall and semantic relevance without retraining \ToolX. $\alpha$ will be 50 in the remaining sections.
\section{Experiments}
\label{sec:results}

% Main results - Tables
\begin{table*}[t]
\centering
\small
\caption{Offline evaluation results of \ToolX~ and baselines. Note that we evaluate recall and semantic relevance of these models on testing user data. For recall, we report recall@K with $K\in\{5,20,100,500\}$. For semantic relevance, we report the average L1/L2 topic cateogory match rate between <trigger,candidate> item pairs. L1 topic divides items into relatively coarse-grained semantic classes and L2 topic divides items into more fine-grained semantic classes. \draft{Note that both MoL and HLLM are modified versions of prior works for item-to-item retrieval; for HSTU, we utilize its item embeddings for item-to-item retrieval in evaluation.}}
\label{tab:main}
\begin{tabular}{p{0.8in}|p{0.8in}p{0.8in}p{0.8in}p{0.85in}|p{0.9in}p{0.9in}}
\toprule
\multirow{2}{*}{Baseline} & \multicolumn{4}{c}{Recall} & \multicolumn{2}{c}{Semantic Relevance}\\ \cline{2-7}
 & Recall@5 & Recall@20 & Recall@100 & Recall@500 & L1 Topic & L2 Topic \\
\midrule
ItemCF~\cite{khoshneshin2010collaborative}    & 1.01\%      & 2.50\%      
                                   & 6.89\%      & 14.88\%    
                                   & 21.31\%     & 17.90\%  \\
NeuCF~\cite{he2017neural}          & 1.10\% (+8.9\%)        & 2.73\% (+9.2\%)     
                                   & 7.51\% (+9.0\%)        & 16.33\% (+9.7\%)    
                                   & 25.77\% (+20.9\%)      & 22.40\% (25.1\%) \\
MoL~\cite{zhai2023revisiting}      & 1.10\% (+8.9\%)        & 2.75\%(+10.0\%)
                                   & 7.49\% (+8.7\%)        & 16.05\%  (+7.9\%) 
                                   & 27.70\% (+30.0\%)      & 24.74\% (+38.2\%)\\
HLLM~\cite{chen2024hllm}           & 1.08\% (+6.9\%)        & 2.67\% (+6.8\%)     
                                   & 7.35\% (+6.7\%)        & 16.07\% (+8.0\%)    
                                   & 28.63\% (+34.4\%)      & 26.25\% (+46.6\%) \\
\draft{HSTU*~\cite{zhai2024actions}}        & 1.05\% (+4.0\%)       & 2.68\% (+7.2\%)     
                                   & 6.46\% (-6.2\%)        & 15.21\% (+2.2\%)    
                                   & 24.56\% (+34.4\%)      & 22.83\% (+46.6\%) \\

\textbf{MTMH (Ours)} & \textbf{1.10\% (+8.9\%)}     & \textbf{2.76\% (+10.4\%)}   
                      & \textbf{7.65\% (+11.0\%)}    & \textbf{17.02\% (+14.4\%)}   
                      & \textbf{30.17\% (+41.6\%)}    & \textbf{28.02\% (+56.5\%)} \\
\bottomrule
\end{tabular}
\end{table*}

\begin{table*}[t]
\centering
\small
\caption{Ablation study of \ToolX. Note that \ToolX-H1 and \ToolX-H2 denote the engagement head and relevance head of \ToolX~respectively. STMH is a mult-head model where each head is trained on single-task learning loss, and MTSH represents a single-head model trained on multi-task learning loss.}
\label{tab:ablation}
\begin{tabular}{p{0.8in}|p{0.8in}p{0.8in}p{0.8in}p{0.8in}|p{0.9in}p{0.9in}}
\toprule
\multirow{2}{*}{Baseline} & \multicolumn{4}{c}{Recall} & \multicolumn{2}{c}{Semantic Relevance}\\ \cline{2-7}
 & Recall@5 & Recall@20 & Recall@100 & Recall@500 & L1 Topic & L2 Topic \\
\midrule
\underline{\ToolX}  & \underline{1.10\%}    & \underline{2.76\%}   
                    & \underline{7.65\%}    & \underline{17.02\%}   
                    & \underline{30.17\%}   & \underline{28.02\%} \\
\ToolX-H1           & 1.11\% (+0.9\%)       & 2.77\% (+0.4\%)     
                    & 7.57\% (-1.0\%)       & 16.16\%  (-5.1\%)   
                    & 28.58\% (-5.3\%)     & 25.73\% (-8.2\%) \\
\ToolX-H2           & 1.12\% (+1.8\%)       & 2.72\% (-1.4\%)    
                    & 6.61\% (-13.6\%)       & 11.37\%  (-33.2\%)
                    & 34.70\% (+15.0\%)      & 33.73\% (+20.4\%) \\
MTSH                & 1.14\% (+3.6\%)       & 2.79\% (+1.1\%)     
                    & 7.09\% (-7.3\%)       & 13.35\%  (-21.6\%)   
                    & 34.15\% (+13.2\%)     & 33.04\% (+17.9\%) \\
STMH                & 1.11\% (+0.9\%)       & 2.75\% (-0.4\%)    
                    & 7.64\% (-1.3\%)       & 14.31\% (-15.9\%)   
                    & 28.97\% (-4.0\%)      & 27.84\% (-0.6\%) \\
\bottomrule
\end{tabular}
\end{table*}

To assess the effectiveness of \ToolX, we conduct a comprehensive evaluation on a commercial recommendation platform serving billions of users. Our evaluation focuses on two key aspects: I2I co-engagement rate (i.e. recall) and I2I semantic relevance.
We start with offline evaluation, where we train \ToolX~alongside other baseline models using user data from Period $T_1$ and then test these models on real user data from Period $T_2$ after $T_1$. The offline evaluation allowed us to compare the performance of \ToolX~with prior SOTAs in a controlled setting (see Section \ref{subsec:main_results}).
To further validate our findings, we deploy \ToolX~on the commercial recommendation platform and compare its performance with the production model. The online evaluation provides valuable insights into how \ToolX~affects the real-world consumption metrics and user-experience-related metrics (see Section \ref{subsec:online}).

\subsection{Baselines and Offline Metrics}
\label{subsec:baselines}
During offline evaluation, we compare \ToolX~with four baseline I2I retrieval models. Note that all baseline models are trained to minimize the same co-engagement loss defined in Eq. (\ref{eq:engage-loss}). Additionally, we employ a  propensity-score-based method to mitigate the selection bias toward popular items during training \cite{chen2023bias}, and a mixed negative sampling strategy in \cite{yang2020mixed} to reduce the selection bias of negative I2I pairs. The baselines differ only from the model architectures and input item features used to generate item embeddings. We describe them in detail below:
\begin{itemize}[leftmargin=*]
    \item ItemCF \cite{khoshneshin2010collaborative}: This model represents each item using an unique embedding vector in Euclidean space purely based on its id, which is originally proposed in \cite{khoshneshin2010collaborative}.
    \item NeuCF \cite{he2017neural}: This model uses a deep neural network (DNN) to learn item embeddings, which has been widely used in prior works. Note that the DNN in this model does not take any item content features as input.
    \item MoL \cite{zhai2023revisiting}: Instead of only taking content-unrelated trigger/item features as input, this model also takes content features of items as DNN input.
    \item HLLM \cite{chen2024hllm}: This model is modified from the most recent work \cite{chen2024hllm}, which takes content embedding generated by pre-trained content encoder models as augmented input of I2I retrieval model.
    \item \draft{HSTU* \cite{zhai2024actions}: This model is one of the SOTA user-to-item retrieval models based on transformers \cite{zhai2024actions}. We use the item embedding learnt by HSTU to perform item-to-item retrieval in evaluation.}
\end{itemize}
% \textit{1) ItemCF \cite{khoshneshin2010collaborative}:} This model represents each item using an unique embedding vector in Euclidean space purely based on its id, which is originally proposed in \cite{khoshneshin2010collaborative}.

% \textit{2) NeuCF \cite{he2017neural}:} This model uses a deep neural network (DNN) to learn item embeddings, which has been widely used in prior works. Note that the DNN in this model does not take any item content features as input.

% \textit{3) MoL \cite{zhai2023revisiting}:} Instead of only taking content-unrelated trigger/item features as input, this model also takes content features of items as DNN input.

% \textit{4) HLLM:} This model is modified from the most recent work \cite{chen2024hllm}, which takes content embedding generated by pre-trained content encoder models as augmented input of I2I retrieval model.

We report both recall and the semantic relevance of \ToolX~ and baselines. Specifically, for recall, we report recall@K on testing data with $K\in\{5,20,100,500\}$. To measure the semantic relevance, we categorize items into distinct topic category groups based on their semantic relevance and report the topic match rate of <trigger,candidate> item pairs.  It's worth noting that L1 topic categories group items into broader topic classes, while L2 topic categories group them into more fine-grained topic classes. We report both to evaluate semantic relevance at varying levels of granularity.
\draft{Our model hyperparameters and training setups are detailed in Table \ref{tab:model_hyperparameters}.}

%%%%%%%%%%%% New %%%%%%%%%%%%%%%%
\begin{table}[htbp]
\caption{Hyperparameters and training details.}
\label{tab:model_hyperparameters}
\centering
\small
\begin{tabular}{p{1.5in}|p{1.5in}}
\toprule
\centering
\textbf{Name} & \textbf{Value} \\
\hline
Number of GPUs & 48 A100s \\
Batch size & 2048 \\
Learning rate & 0.01 \\
Optimizer & Adagrad \\
Training epoch & 1 \\
Item embedding dim & 128 \\
Content embedding dim & 128 \\
Visual encoder model & VIT-H14-632M \\
Text encoder model & XLM-R-Large-550M \\
Visual encoder output dim & 1024 \\
Text encoder output dim & 1024 \\
Fusion MLP output dim & 128 \\
Output dim of MLP in MTMH & 128 \\
\bottomrule
\end{tabular}
\end{table}
%%%%%%%%%%%% New %%%%%%%%%%%%%%%%

\vspace{-3mm}
\subsection{Main Results}
\label{subsec:main_results}
We first present offline evaluation results of \ToolX~and baselines. As shown in Table \ref{tab:main}, \ToolX~achieves both the highest recall and semantic relevance compared with all baselines. Specifically, \ToolX~ increases the recall@500 by at least 4.2\% and up to 14.4\%. In terms of L2 topic relevance, \ToolX~ outperforms all baselines by at least 6.7\% and up to 56.5\%.

It is worth noting that both MoL and HLLM have better I2I semantic relevance but worse recall@500 compared with NeuCF. As mentioned in Section \ref{subsec:baselines}, MoL takes content features as input to generate item embeddings and HLLM uses content embeddings generated by LLM as input, which can greatly improve I2I semantic relevance. However, this usually comes with the expense of sacrificing I2I co-engagement rate. In contrast, \ToolX~ is capable of preserving both highly co-engaged items and semantically relevant items due to its multi-head serving design (see Section \ref{subsec:serving}), achieving both improved recall and semantic relevance.

\subsection{Ablation Study}
\label{subsec:ablation}
Next, we conduct the ablation study of \ToolX~ by comparing it with the following baselines:

% \textit{1) \ToolX-H1}: It only uses the engagement head of \ToolX.

% \textit{2) \ToolX-H2}: It only uses the relevance head of \ToolX.

% \textit{3) MTSH (Multi-task single-head)}: This is a single-head model trained by minimizing multi-task learning loss (see Eq. (\ref{eq:loss_mt})).

% \textit{4) STMH (Single-task multi-head)}: This is a multi-head model with one engagement head and one relevance head. Different from \ToolX, the engagement head is trained to solely minimize the co-engagement loss (see Eq. (\ref{eq:engage-loss})), while the relevance head is trained to solely minimize relevance loss (see Eq. (\ref{eq:relev-loss})).

\begin{itemize}[leftmargin=*]
    \item \ToolX-H1: It only uses the engagement head of \ToolX.
    \item \ToolX-H2: It only uses the relevance head of \ToolX.
    \item MTSH (Multi-task single-head): This is a single-head model trained by minimizing multi-task learning loss (see Eq. (\ref{eq:loss_mt})).
    \item STMH (Single-task multi-head): This is a multi-head model with one engagement head and one relevance head. Different from \ToolX, the engagement head is trained to solely minimize the co-engagement loss (see Eq. (\ref{eq:engage-loss})), while the relevance head is trained to solely minimize relevance loss (see Eq. (\ref{eq:relev-loss})).
\end{itemize}

As shown in Table \ref{tab:ablation}, \ToolX-H1 (i.e. the engagement head) has the best recall compared with other baselines, and \ToolX-H2 (i.e. the relevance head) has the highest I2I semantic relevance while the lowest recall@500 compared with other models. By merging candidates retrieved from both heads, the recall@500 of \ToolX~ is further increased by 5.3\% on top of \ToolX-H1. At the same time, it improves I2I semantic relevance by 5.6\% w.r.t. L1 and 8.9\% w.r.t. L2 compared with \ToolX-H1, since it preserves both co-engaged and semantically relevant candidates retrieved by \ToolX-H2.

Moreover, we observe that both STMH and MTSH trade recall for better semantic relevance. For instance, compared with \ToolX-H1 and baselines in Table \ref{tab:main}, they exhibit significantly better semantic relevance but much worse recall. By contrast, \ToolX is the only model which can improve I2I semantic relevance without trading recall, which demonstrates the effectiveness of multi-head modeling and serving strategy in Section \ref{sec:method}.

\subsection{Recall and Relevance Trade-off}
\label{subsec:tradeoff}
\begin{figure}[t]
    \centering
    \begin{subfigure}{0.48\linewidth}
        \includegraphics[width=\linewidth]{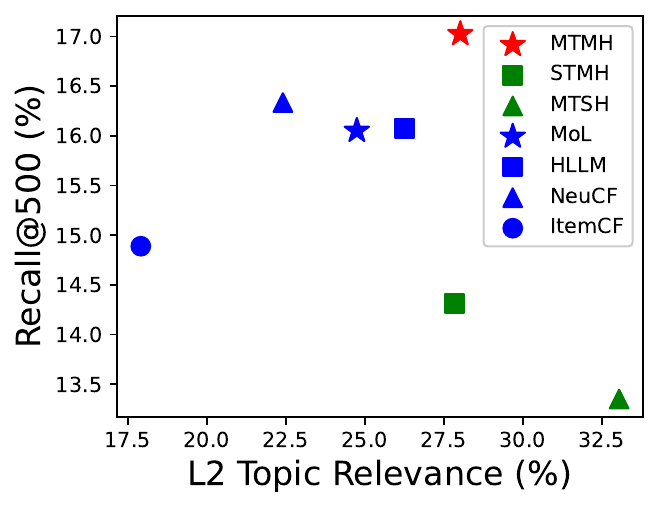}
        \caption{\ToolX~vs baselines.}
        \label{fig:trade-all}
    \end{subfigure}
    \begin{subfigure}{0.48\linewidth}
        \includegraphics[width=\linewidth]{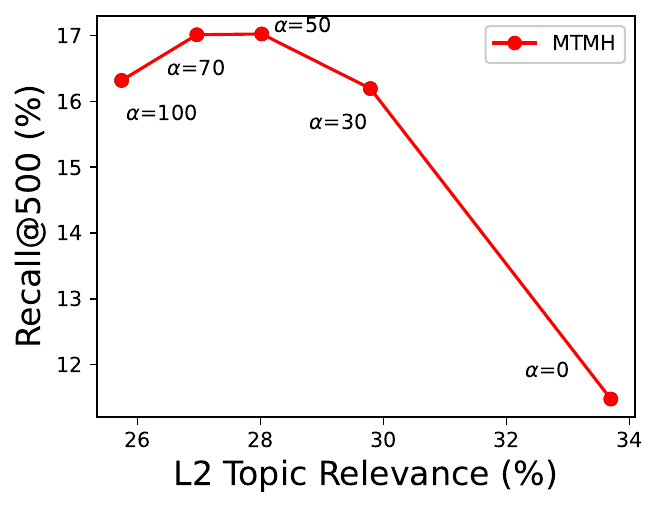}
        \caption{\ToolX~with varying $\alpha$.}
        \label{fig:trade-mtmh}
    \end{subfigure}
    \caption{Recall and semantic relevance trade-off evaluation results with varying $\alpha$. Note that top right part of these figures represents both high recall and semantic relevance. $\alpha$ in \ToolX~ controls the percentage of candidates from its engagement head. In the left figure, the default $\alpha$ value 50 is used.}
    \label{fig:tradeoff}
    \vspace{-3mm}
\end{figure}

In this subsection, we evaluate the recall and semantic relevance trade-off performance of baselines and \ToolX~ with varying $\alpha$ and \draft{$w_r$}. Note that $\alpha$ is a serving hyperparameter controlling the quota for the engagement head during serving time, \draft{while $w_r$ is a training hyperparameter determining the weight of relevance loss (see Eq. \ref{eq:loss_mt})}. We use recall@500 and L2 topic relevance as our recall and semantic relevance metrics, and report the results in Figure \ref{fig:tradeoff}.

\noindent\textbf{Varying $\alpha$.} As shown in Figure \ref{fig:trade-all}, \ToolX~ is able to achieve both high recall and semantic relevance, outperforming all baselines except MTSH (multi-task single-head model) in terms of both metrics. While MTSH achieves the best L2 topic relevance, its recall@500 is significantly lower than that of the other models. Figure \ref{fig:trade-mtmh} demonstrates how varying serving parameter $\alpha$ can affect the recall and semantic relevance trade-off of \ToolX. We observe that decreasing $\alpha$ improves the I2I semantic relevance, since more semantically relevant items retrieved by the relevance head of \ToolX~ are preserved. By contrast, increasing $\alpha$ boosts the recall, by selecting more highly co-engaged items retrieved by the engagement head of \ToolX. Moreover, we observe that when $\alpha$ is larger than 50, decreasing $\alpha$ can enhance I2I recall and semantic relevance simultaneously. This is expected since items with less co-engagement efficiency retrieved by the engagement head of \ToolX~ are replaced by highly relevant and co-engaged items retrieved from the relevance head of \ToolX~. However, keeping reducing $\alpha$ leads to the drop of recall, specifically when $\alpha > 70$, since the relevance head may ignore highly co-engaged but less semantically relevant items (e.g. popular items). In summary, we observe that $\alpha=50$ provides us with the optimal recall and semantic relevance trade-off.

Note that in practice, $\alpha$ can be flexibly adjusted to adapt to the specific requirements in production. For instance, for some applications where I2I semantic relevance is more important, smaller $\alpha$ can be used; while for applications focusing on more co-engagement rate, larger $\alpha$ should be used. By enabling $\alpha$ as a serving parameter, \ToolX~model can be deployed to serve different purposes without being retrained.

\noindent\textbf{Varying $w_r$.} \draft{Table \ref{tab:wr} shows the additional results with varying values of $w_r$ in the table below. It can be seen that larger $w_r$ generally increases relevance but decreases recall. Overall, we observe that $w_r$=0.5 provides a decent trade-off between recall and relevance. Therefore, we use 0.5 as a default value for $w_r$ during both offline and online experiments.}

\begin{table}[h]
  \caption{Recall and semantic relevance trade-off evaluation results with varying $w_r$.}
  \label{tab:wr}
  \centering
  \begin{tabular}{p{0.8in}|p{0.8in}|p{0.6in}p{0.6in}}
    \toprule
    \multirow{2}{*}{$w_r$} & \multicolumn{1}{c|}{Recall} & \multicolumn{2}{c}{Semantic Relevance}\\ \cline{2-4}
     & $\Delta$Recall@500 & $\Delta$L1 Topic & $\Delta$L2 Topic\\
    \hline
    0 & -2.99\% & -13.57\% & -19.87\% \\
    \hline
    0.25 & +0.96\% & -3.81\% & -5.54\% \\
    \hline
    \textbf{0.5 (Default)} & -- & -- & -- \\
    \hline
    1.0 & -0.09\% & +3.07\% & +4.45\% \\
    \hline
    5.0 & -7.38\% & +9.01\% & +13.22\% \\
    \bottomrule
  \end{tabular}

\end{table}

\subsection{Embedding Convergence of Fresh Content}
\begin{figure}[t]
    \centering
    \includegraphics[width=0.7\linewidth]{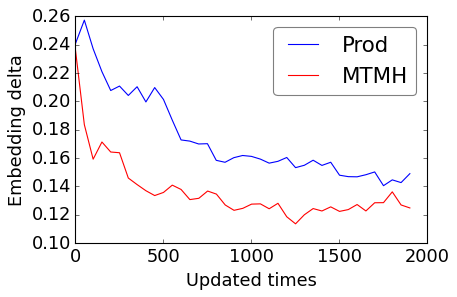}
    \vspace{-3mm}
    \caption{MTMH embedding convergence of fresh content.}
    \label{fig:convergence}
    \vspace{-5mm}
\end{figure}

% Most industry retrieval models including I2I are trained heavily on user engagement data, resulting in severe bias towards old contents. However, one important aspect in recommendation system is the fast delivery of fresh contents (w/ little user engagement), which is critical of enhancing the overall user experience. To achieve this, the I2I retrieval model needs to learn the embeddings of fresh content faster. This means that the embedding of fresh content should converge quickly during model training. To investigate whether \ToolX~can accelerate fresh content delivery, we analyze the convergence speed of fresh content embeddings during \ToolX~training. Figure \ref{fig:convergence} compares the fresh content embedding convergence speed of \ToolX~ with our production model, where the $x$-axis presents the number of updated times for the fresh content embeddings and the $y$-axis represents the embedding delta. Note that the embedding delta at step $t$ is defined as the L2 distance between embeddings at step $t$ and embeddings at step $T=2000$. We observe \ToolX~ achieves faster embedding convergence for fresh content compared to the production model. This finding is further validated by our online experimental results in Section \ref{subsec:online}, which show that \ToolX~ successfully delivers more fresh content, aligning with our expectations based on its improved embedding convergence.
Most industry retrieval models, including I2I, rely heavily on user engagement data, leading to a strong bias toward older content. However, a key aspect of recommendation systems is the rapid delivery of fresh content, even with minimal user engagement, as it enhances the overall user experience. To achieve this, the I2I retrieval model must learn fresh content embeddings more quickly, ensuring their fast convergence during training.
To assess whether \ToolX~ accelerates fresh content delivery, we analyze the convergence speed of fresh content embeddings during \ToolX~ training. Figure \ref{fig:convergence} compares the embedding convergence speed of fresh content between \ToolX~ and our production model. The $x$-axis represents the number of updates to fresh content embeddings, while the $y$-axis denotes the embedding delta—defined as the L2 distance between embeddings at step $t$ and those at step $T=2000$. Our results show that \ToolX~ achieves faster embedding convergence for fresh content compared to the production model. This finding is further validated by online experiments (Section \ref{subsec:online}), which confirm that \ToolX~ successfully delivers more fresh content, aligning with expectations based on its improved embedding convergence. The reason is that our MTMH not only learns the item embeddings from the co-engagement data, but also incorporates the content semantic relevance through the multi-task learning, improving the model generalization on fresh content.

\subsection{Online A/B Testing}
\label{subsec:online}
We deployed \ToolX~on a real-world platform serving billions of users and conducted a 7-day evaluation to assess its effectiveness. We observe significant gains on both topline consumption metrics and user-experience-related metrics. Table \ref{tab:online} reports the improvements in several key metrics, which are described below.\\
\noindent\textbf{Daily active users (DAU).} This metric measures the number of unique users who engage with the platform on a daily basis. We observe 0.05\% increase in DAU, indicating that more users are using the platform each day.\\
\noindent\textbf{Daily time Spent.} This metric measures the amount of time users spent on the platform within a day. The deployment of \ToolX~ leads to a 0.22\% increase in users' time spent, suggesting that users are more engaged with the recommended items on the platform.\\
\noindent\textbf{Distinct item views.} This metric counts the number of unique items viewed by users on the platform in a day. An increase in distinct item views indicates that more diverse and unique items are retrieved and recommended to users. We report that \ToolX~ brings 0.31\% more distinct items into the platform. \\
\noindent\textbf{Percentage of fresh content.} This metric measures the percentage of fresh content with age less than 48 hours on the platform. \ToolX~ improves this metric by 0.25\%, indicating that users are seeing more fresh content.\\
\noindent\textbf{Novel interest discovery rate.} This metric tracks the number of new interests or topics that users discover on the platform. An increase in novel interest discovery rate indicates that users' new areas of interest can be discovered by the recommendation system faster. We observe that \ToolX~ can also increase this metric by 0.33\%.\\
\noindent\textbf{User interest recall.} This metric measures how well the platform is able to recommend content that aligns with a user's existing interests. An improvement in user interest recall suggests that our model is able to retrieve more semantically relevant content. As expected, \ToolX~ successfully moves this metric up by 0.14\%.

In summary, we conclude that \ToolX~ improves both I2I co-engagement efficiency and semantic relevance during online A/B testing, consistent with what we observe during offline evaluation.

% \subsection{Case Study}
% \textcolor{red}{[Pending: We can show some I2I retrieval examples here. But need to remove sensitive parts in videos.]}

\section{Related Work}
\label{sec:related}
\noindent\textbf{I2I retrieval models.} Early I2I retrieval methods include collaborative filtering \cite{rendle2012bpr,sarwar2001item,su2009survey}, matrix factorization \cite{hu2008collaborative,Koren2008FactorizationMT,koren2009matrix} and neighbor-based methods \cite{linden2003amazon,Sarwar2000AnalysisOR}. They primarily focus on modeling interactions between raw item IDs without considering rich features like item attributes or content features \cite{Khoshneshin2010CollaborativeFV,liang2016modeling,rendle2010factorization,xin2018batch}, which are insufficient for capturing complex patterns and relationships between items. Recent years, deep neural networks (DNNs) have been used in I2I retrieval to capture complex patterns and relationships between items. Among them, two-tower model architecture has emerged as a dominant paradigm, offering both effectiveness and efficiency \cite{covington2016deep,huang2013learning,yu2021dual}. Along this line of work, various techniques like adaptive mechanisms \cite{Li2022IntTowerTN,xu2022mixture,yu2021dual} and self-attention \cite{li2019multi,xiao2024deep} have been used to enhance two-tower models with richer input features while maintaining computational efficiency. However, these models are trained purely based on co-engagement data without considering I2I semantic relevance. To tackle the semantic understanding challenge, various approaches have been proposed to take content features as model input to improve I2I semantic relevance \cite{li2019multi,liu2020octopus,tan2021sparse,hidasi2013context,lv2019interest,wang2022surrogate,Hidasi2017RecurrentNN,Lops2011ContentbasedRS}. Different from prior works, \ToolX~ provides a principled approach for jointly optimizing co-engagement efficiency and semantic relevance via multi-task learning loss without taking content features as input.

\noindent\textbf{Large foundation models for retrieval.} With the emergence of large foundation models (e.g. large language models (LLMs)), there has been growing interest in leveraging their superior semantic understanding capabilities for recommendation tasks~\cite{Lin2023HowCR,Zhang2025ColdStartRT,Zhao2023ASO,Wu2023ASO,Fan2023RecommenderSI,Li2023LargeLM,Boz2024ImprovingSR,Wang2024TowardsNL,Liu2024LargeLM}. One line of work use LLMs for generative recommendations through prompt engineering~\cite{dai2023uncovering,Geng2022RecommendationAL}. Another line of work integrate LLMs into retrieval systems \cite{chen2024hllm,dai2023uncovering,li2023pbnr,wang2023recmind,Wang2024RethinkingLL} to better understand complex item relationships. However, these approaches face practical deployment challenges due to hugh runtime cost. More recent works have explored methods to inject LLM knowledge into recommendation models~\cite{Chu2023LeveragingLL,Wang2023FLIPFA}, either by enhancing item representations through content feature extraction~\cite{Wei2023LLMRecLL,Xi2023TowardsOR,Zheng2024HarnessingLL}, or leveraging LLMs for data augmentation~\cite{Chen2024EnhancingIR} and knowledge distillation~\cite{Xu2024ASO}. 
In contrast to prior work, \ToolX~ provides a principled approach for distilling LLMs' knowledge into retrieval models via multi-task learning without increasing model complexity. Moreover, the multi-head architecture design in \ToolX enables us to flexibly trade between co-engagement efficiency and semantic relevance, which is unexplored in prior works.
\begin{table}[t]
\centering
\small
\caption{Online A/B testing results for \ToolX~.}
\label{tab:online}
\begin{tabular}{p{2.0in}p{0.6in}}
\toprule
Metrics & Changes \\
\midrule
Daily active users & +0.05\% \\
Daily time spent & +0.22\% \\
Distinct item views & +0.31\% \\
Perentage of fresh content & +0.25\% \\
Novel interest discovery rate & +0.33\% \\
User interest recall & +0.14\% \\
\bottomrule
\end{tabular}
\vspace{-5mm}
\end{table}
\section{Conclusion}
\label{sec:conclusion}

This paper proposes \ToolX, a multi-task and multi-head item-to-item (I2I) retrieval model that addresses the fundamental trade-off between recall and semantic relevance. \ToolX~provides a principled approach for jointly optimizing I2I co-engagement rate and semantic relevance, via a multi-task learning loss and a multi-head retrieval architecture. Our offline experimental results demonstrate that \ToolX~improves I2I retrieval recall by up to 14.4\% and semantic relevance by up to 56.6\%, outperforming all baselines. Our online A/B testing further verifies its effectiveness in enhancing both topline consumption metrics (e.g. daily active user and time spent) and user-experience-related metrics (e.g user interest recall, novel interest discovery rate, content diversity, and freshness). Overall, this work has the potential to significantly enhance the performance of recommendation systems in various applications.

\bibliographystyle{ACM-Reference-Format}
% \balance
\bibliography{main}

\end{document}